\documentstyle[aps,pra,multicol,psfig]{revtex}
\begin{document}

%\title{Disorder dominated critical behavior of the random Potts model}
%\title{Exact critical exponents of the 2d random Potts model in the large-$q$ limit}
\title{Phase transition in the 2d random Potts model in the large-$q$ limit}

\author{J-Ch. Angl\`es d'Auriac$^{1}$ and
F. Igl\'oi$^{2,3}$}

%\affiliation{
\address{
$^1$ CNRS-CRTBT B. P. 166, F-38042 Grenoble, France\\
$^2$ Research Institute for Solid State Physics and Optics, 
H-1525 Budapest, P.O.Box 49, Hungary\\
$^3$ Institute of Theoretical Physics,
Szeged University, H-6720 Szeged, Hungary
}

\date{\today}

\maketitle

\begin{abstract}
Phase transition in the two-dimensional $q$-state Potts model
with random ferromagnetic couplings in the large-$q$ limit is
conjectured to be described by the isotropic version of the infinite
randomness fixed point of the random transverse-field Ising spin
chain. This is supported by extensive numerical studies with a
combinatorial optimization algorithm giving estimates for the
critical exponents in accordance with the conjectured values:
$\beta=(3-\sqrt{5})/4$, $\beta_s=1/2$ and $\nu=1$. The specific
heat has a logarithmic singularity, but at the transition point there are
very strong sample-to-sample fluctuations. Discretized randomness results
in discontinuities in the internal energy.
\end{abstract}

%\maketitle

\newcommand{\bc}{\begin{center}}
\newcommand{\ec}{\end{center}}
\newcommand{\be}{\begin{equation}}
\newcommand{\ee}{\end{equation}}
\newcommand{\beqn}{\begin{eqnarray}}
\newcommand{\eeqn}{\end{eqnarray}}

\begin{multicols}{2}
\narrowtext

Quenched disorder could often change the properties of phase transitions
but exact information is scarce about the singularities in random fixed points,
in particular about isotropic classical systems with short range interactions.
For most of random classical systems - such as in spin glasses and random ferromagnets -
the critical behavior is governed by conventional random fixed
points (CRFP-s), in which the strength of disorder remains finite under renormalization.

There is, however, a class of strongly anisotropic two-dimensional (2d)
random models, in which the disorder is strictly
correlated in 1d and there are exact results about
their critical behavior. These models, such as
the McCoy-Wu model\cite{mccoywu}, are closely related to random quantum spin chains,
and their critical behavior is
usually governed by an infinite randomness fixed point\cite{2drg} (IRFP), in which - under
renormalization - the strength of disorder growths without limits.
In an IRFP the singularities are primary determined by disorder effects and for almost
every studied, strongly disordered, 1d quantum (or 2d classical) systems the critical exponents
are the same as in the universality class of the random
transverse-field Ising model\cite{fisher} (RTIM).

Having in mind the known relation between the critical behavior of 2d classical and
corresponding 1d quantum systems\cite{kogut} with non-random couplings one might ask the
question if a similar correspondence exists for random systems and in particular what is the
2d isotropic classical counterpart of the RTIM. Some hints about a possible 2d isomorphism
of the RTIM is presented in Ref.\onlinecite{profile} in which the operator profiles of the RTIM
are found to be conformally invariant, the property of which should be shared with
some 2d isotropic random system, which - by definition - could obey conformal
symmetries\cite{cardy}.

In this Letter we conjecture that the possible candidate for this r\^ole is 
the $q$-state Potts model\cite{Wu} with random ferromagnetic bonds (random bond Potts model - RBPM)
in the large-$q$ limit and present the following arguments. First, we note that in the above
limit the high-temperature expansion of the
RBPM is dominated by a single diagram\cite{JRI01}, $\cal F$, so that the critical behavior is primary
determined by disorder effects as in the IRFP of the RTIM. Second, we have performed extensive
numerical calculations, in which $\cal F$ is exactly determined by a very efficient combinatorial
optimization algorithm\cite{aips02} and the obtained bulk and surface magnetization scaling dimensions,
$x$ and $x_s$, respectively, are found in good agreement with the corresponding exact values for
the RTIM\cite{fisher}:
\be
x=\frac{3-\sqrt{5}}{4},\quad x_s=\frac{1}{2}\;.
\label{exponents}
\ee
Third, we point out on apparent topological similarities between the ground state of
the RTIM and the fractal structure of the optimal graph, $\cal F$, which could then explain
the isomorphism between the two problems.

In the following we introduce the RBPM by the Hamiltonian:
\be
-\frac{H}{kT}=\sum_{\langle ij \rangle} K_{ij} \delta(\sigma_i,\sigma_j)\;,
\label{hamilton}
\ee
where the spin variable at site $i$ is $\sigma_i=0,1,\dots,q-1$, $K_{ij}>0$
are random ferromagnetic couplings and the summation runs over nearest neighbor
pairs. In the random cluster representation the partition function of the system
is expressed in terms of $v_{ij}=\exp K_{ij}-1$ as\cite{JRI01}
\be
Z=\sum_F q^{C(F)} \prod_{ij \in F} v_{ij}\;,
\label{Z}
\ee
where the summation runs over all subset of bonds, $F$, and $C(F)$ is the number of
connected components of $F$, counting also the isolated sites. Having the
parameterization, $v_{ij}=q^{\alpha_{ij}}$ the partition function is expressed as
\be
Z=\sum_F q^{f(F)},\quad f(F)=C(F)+ \sum_{ij \in F} \alpha_{ij}\;,
\label{Z-q}
\ee
which in the large-$q$ limit is indeed dominated by the largest contribution,
$f^*={\rm max}_F f(F)$, and the partition function is
asymptotically given by $Z=N q^{f^*}$ where the number of optimal sets (OS-s), $N$, is
likely to be one.

One of the few rigorous results about random systems is due to Aizenman and Wehr\cite{aizenmanwehr}
who states that in the 2d RBPM the internal energy is continuous at the phase transition point for
any $q$, if the probability distribution of the couplings is absolutely continuous.
In numerical calculations\cite{pottsmc,pottstm} one usually studies the properties of the
system at the phase transition point, which is known by duality\cite{dom_kinz}, and
universal, i.e. disorder independent, critical behavior
has been observed even for atomistic (c.f. bimodal) distributions. 
The scaling dimension, $x$, is a monotonously increasing function of $q$, but its saturation value
in the large-$q$ limit is difficult to be estimated due to strong
logarithmic corrections\cite{jacobsenpicco}.
The correlation length exponent is close to $\nu=1$, for any
value of $q$. Note, however, that by transfer matrix calculation Jacobsen and Cardy\cite{pottstm}
have obtained $\nu<1$ for the bimodal distribution of disorder, which violates the rigorous
upper bound, $\nu \ge 2/d$\cite{ccfs}.

In the large-$q$ limit the critical properties of the RBPM are related to the structure
of the clusters in the OS in close analogy with percolation\cite{staufferaharony}.
In the paramagnetic phase there are only finite clusters in the OS
and the linear extent of the largest clusters is used to define the correlation length, $\xi$.
In the ferromagnetic phase there is an infinite cluster and the ratio of lattice points
belonging to it is related to the (finite) magnetization of the RBPM.
Finally, at the phase transition point the largest (infinite) cluster is a fractal, and its
fractal dimension, $d_f$, is related to the magnetization scaling dimension as $d=d_f+x$.
Similarly, the sum of the surface
fractal dimension of the percolating cluster, $d_f^s$, and the anomalous dimension of the
surface magnetization, $x_s$, gives the Euclidean dimension of the surface:
$d^s=1=d_f^s+x_s$.

In Ref.\onlinecite{JRI01} the OS was approximatively calculated by the simulated annealing
method. Results, obtained on relatively small lattices (up to $24
\times 24$), such as $x=0.17-0.19$ were consistent with the conjectured value in
Eq.(\ref{exponents}), however, due to
the relatively large error and also due to the lack of precise estimates about another
exponents we could not make a definite statement about the universality class
of the transition.

In the present Letter we apply a recently developed combinatorial optimization
algorithm\cite{aips02} with which we can determine the {\it exact} OS in strongly
polynomial time, i.e. the time of computation does not depend on the form of the
disorder. With this algorithm we could treat far larger systems as before,
averages and distributions were calculated for systems with $L=32,~64,~128$ and $256$
over at least thousand disorder realizations.
(We have also analyzed a few samples of $512 \times 512$).
As a consequence our estimates of the critical exponents become much more accurate than before
and we studied, at the first time, also the surface properties of the RBPM.

In most of the calculation the couplings and thus the parameters, $\alpha_{ij}$,
were taken from a bimodal distribution:
$P(\alpha)=[\delta(w+\Delta w - \alpha) + \delta(w-\Delta w - \alpha)]/2$
with $w > \Delta w > 0$, so that the distance of the critical temperature, $t$,
is measured by $t \approx 1 - 2w$. Having the parameter $\Delta w=1/3$
the microscopic length-scale in the problem\cite{JRI01},
$l_c \approx (2 \Delta w)^{-2}$, was not too large. We also used the continuous
(uniform) distribution: $P_u(\alpha)=1/(2 u)$, for $0 \le \alpha < u$ and zero otherwise. 
Here the distance from the critical point
is given by $t \approx 1-u$. To calculate bulk (surface) quantities
we applied periodic (open) boundary conditions (b.c.-s).

%%%%%%%%%%%%%%%%%%%%%%%%%%%%%%%% FIG. 1 %%%%%%%%%%%%%%%%%%%%%%%%%%%%%%%%%%%%%%
\begin{figure}[b]
\centerline{\psfig{file=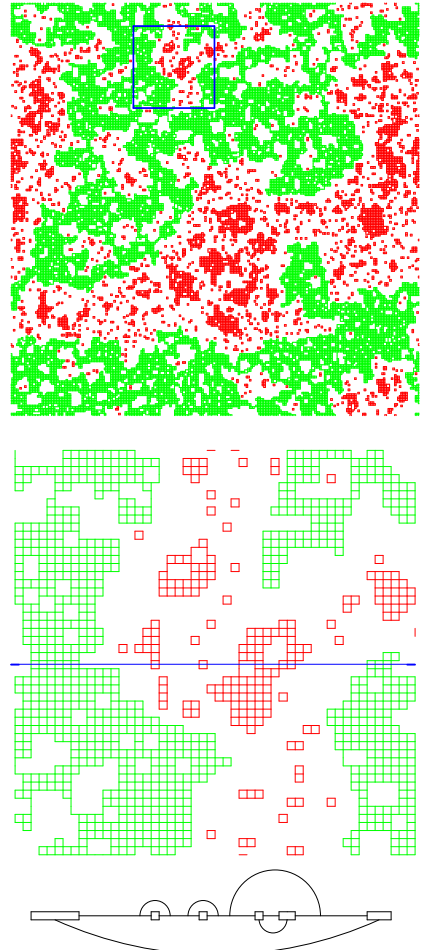,height=12.75cm}}
\vskip 0.2truecm
\caption{top:  OS for a typical disorder realization at the critical point on
a $256 \times 256$ lattice with periodic boundary conditions. Percolating and finite clusters
are marked with gray (green) and dark (red) lines, respectively. middle:  Enlargement of a square
proportion of size $48 \times 48$ in the upper-middle part of the OS to illustrate
self-similarity. bottom: The connectivity structure of the OS along a line, which
consists of six connected units (``spins'') and five open units (``bonds'').}
\label{FIG01}
\end{figure}
%%%%%%%%%%%%%%%%%%%%%%%%%%%%%%%%%%%%%%%%%%%%%%%%%%%%%%%%%%%%%%%%%%%%%%%%%%%%%

First, we considered the behavior of the system at the transition point, when the cluster
structure of a typical OS is shown in Fig. \ref{FIG01}. As illustrated in the middle
of Fig. \ref{FIG01} the OS is self-similar and its topology, being isotropic,
can be conveniently represented by the connectivity structure (CS) of the OS
at a given line, as shown in the bottom of Fig. \ref{FIG01}. We are going to analyze
the CS after presenting the numerical results.

\begin{figure}[b]
\centerline{\psfig{file=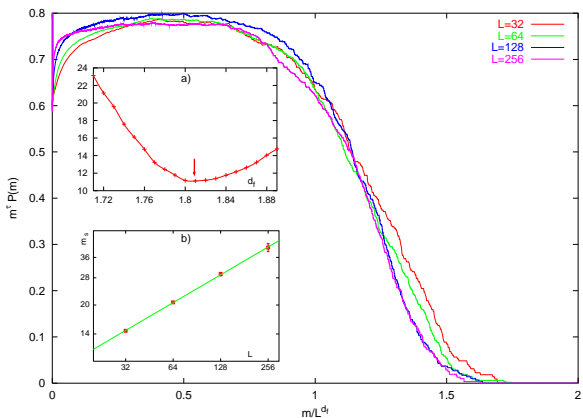,height=5.75cm,angle=-90}}
\vskip 0.2truecm
\caption{Scaling collapse of the reduced probability distribution function, $R(m,L) m^{\tau}$,
in Eq.(\ref{distr_m}) with $d_f=(5+\sqrt{5})/4=1.809$. Inset a): Surface of the
collapse region for different values of $d_f$. The arrow shows the conjectured value.
b) Average mass of surface sites of the largest cluster vs. linear size of the system
in a log-log plot. The slope of the straight line corresponds to $d_f^s=0.495(10)$.}
\label{FIG02}
\end{figure}

We have checked that asymptotically half of the sites of the OS are
isolated: the probability
$P(L)$ that a given site belongs to a cluster having two or more sites, behaves like 
$P(L) \simeq  1/2 \left( 1 + 1/(2\ln L) \right)$.
We have also checked that the largest
connected cluster is indeed a fractal (and not a multi-fractal). Its fractal dimension, $d_f$, is calculated
by direct dimensional analysis and by finite-size scaling of the average mass of the largest
cluster. Here we present an analysis of the probability distribution function,
$R(m,L)$, which measures the fraction of clusters having a size at least $m$ and which,
according to scaling theory\cite{staufferaharony}, should asymptotically behave as:
\be
R(m,L)=m^{-\tau} \tilde{R}(m/L^{d_f})\;,
\label{distr_m}
\ee
with $\tau=(2-d_f)/d_f$. With the conjectured value of $d_f=2-x=(5+\sqrt{5})/4=1.809$
we obtained a very good scaling collapse, which is shown in Fig.\ref{FIG02}.

To characterize the accuracy of the collapse we have
measured the surface of the overlap of the scaled curves with varying value of $d_f$.
As shown in the inset a) to Fig. \ref{FIG02} the best collapse is indeed around the
conjectured value, so that our estimate $x=0.190(5)$ has a relatively small error.

%%%%%%%%%%%%%%%%%%%%%%%%%%%%%%%% FIG. 2 %%%%%%%%%%%%%%%%%%%%%%%%%%%%%%%%%%%%%%
%%%%%%%%%%%%%%%%%%%%%%%%%%%%%%%%%%%%%%%%%%%%%%%%%%%%%%%%%%%%%%%%%%%%%%%%%%%%%

Next, we considered the surface magnetization properties of the model by
using open b.c.-s and measuring the average mass of surface sites of the largest cluster, $m_s$.
As shown in the inset b) to Fig. \ref{FIG02} it has an asymptotic size dependence
$m_s \sim L^{d_f^s}$, with $d_f^s=0.495(10)$, which is in very good agreement with the
conjectured value of $d_f^s=1-x_s=1/2$, see Eq.(\ref{exponents}).

In the following we analyze in more details the fractal structure of the OS and point out
the topological similarities with the ground state wave function of the RTIM, which could
explain the appearance of the same critical exponents in the two problems. As already noted
in the bottom of Fig.\ref{FIG01} the CS consists of connected units (CU-s)
(corresponding to spins in the RTIM) of variable length, $l_s$, and moments, $\mu$, and of
open units (OU-s) (corresponding to bonds in the RTIM) of variable
length, $l_b$. If two neighboring CU-s, with parameters  $l_s^1$, $\mu^1$ and $l_s^2$, $\mu^2$,
separated with an OU of $l_b$, belong to the same cluster, it is merged to an effective
CU (represented by a connecting line in Fig. \ref{FIG01}), with length
$l_s^{12}=l_s^1+l_s^2+l_b$ and moment $\mu^{12}=\mu^1+\mu^2$. This is precisely
equivalent of a strong bond decimation in the RTIM, which is one ingredient of the
strong disorder renormalization group (SDRG) method\cite{mdh,fisher}. Similarly, if a CU
with $l_s$ and $\mu$ and with neighboring bonds of lengths: $l_b^1$ and $l_b^1$, is isolated,
it does not contribute to any larger cluster, therefore - at larger length-scales - can be
eliminated (represented by an overgoing line in Fig. \ref{FIG01}) and the new effective
bond has a length of $l_b^{12}=l_b^1+l_b^2+l_s$. This process then equivalent of a strong
field decimation in the SDRG for the RTIM. Thus we can conclude that for any CS of the OS
in the RBPM one can construct an equivalent ground state of the RTIM,
and one can give a set
of couplings, $J_i$, and transverse fields,$h_i$, for which the given ground state is realized.
We recall that in the RTIM the bulk (surface) magnetization is related to the
average moment of a bulk (surface) effective spin\cite{fisher} in close analogy with the
computation of the same quantities in the OS of the RBPM.
If we now assume that in the two problems the statistics of the appearance of
states with equivalent topology is also similar, we arrive to the conjectured
relation about the critical exponents in Eq.(\ref{exponents}).

\begin{figure}[b]
\centerline{\psfig{file=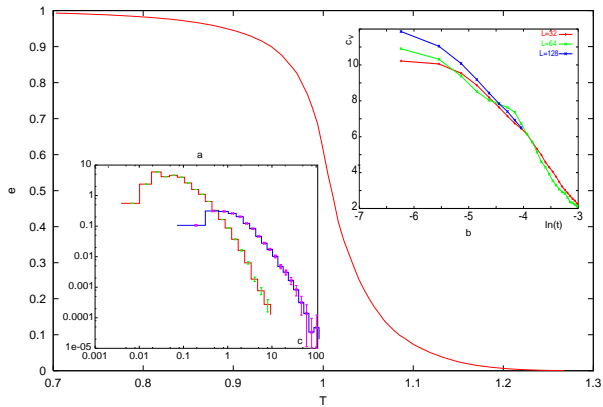,height=5.75cm}}
\vskip 0.2truecm
\caption{Average internal energy for the continuous
distribution. Inset a): distribution of the finite-size latent heat
at the right side of the transition point for the bimodal
distribution for $L=32$ (left) and $L=128$ (right). Inset b): the
average specific heat for the continuous
distribution as a function of $\ln |t|$ in the region $Lt \gg 1$.}
\label{FIG03}
\end{figure}

In the following we turn to study the energy-density of the model outside the transition point.
To do this one should determine the OS at different temperatures,
however, in a finite system of size, $L$, there are only a finite
number of different OS-s, their typical
number being $L$, independently of the type of disorder. This result
can be interpreted that two neighboring OS-s differ in average by one
line ($\sim L$) of edges. For a given sample
the free energy is a piece-wise linear function of $t$ and the
internal energy is a step-like function having typically
$L$ steps. Averaging over disorder the average internal energy
becomes continuous for continuous distributions (see. Fig.\ref{FIG03}), whereas for discrete
distributions some discontinuities, located at special isolated points,
remain. (The discontinuous behavior of the internal energy in this
case is somewhat analogous to that of the magnetization of the random-field Ising model at $T=0$
having the same bimodal distribution\cite{RFIM}.)
There is a discontinuity at the phase-transition (self-dual) point, as we
illustrate with the sequence of finite-size latent heats:
$[\Delta E(32)]_{\rm av}=0.0456(90)$, $[\Delta E(64)]_{\rm av}=0.0477(52)$,
$[\Delta E(128)]_{\rm av}=0.0484(28)$ and $[\Delta E(256)]_{\rm av}=0.0474(14)$
for the bimodal distribution with $\Delta w=1/3$, which approach a
finite value in the thermodynamic limit. The discontinuities in the
internal energy are due to degeneracies, which are connected also to finite
clusters. For example at the self-dual point any lattice point which
has two couplings - one weak and one strong - to an existing cluster, could
either be connected or disconnected. Thus the jumps in the
internal energy are related not exclusively to the largest cluster.
The contribution of the largest cluster only, which has a diverging size $\xi$,
is expected to cause singularity in the specific heat
at the two sides of the transition point.

This true singularity of the
specific heat, which is expected to be independent of the type of
disorder, will be investigated numerically in the following.
Here, as usual, the thermodynamic limit should be taken
first and then approach the transition point.
In our numerical study we observed that close to the transition point,
in the order of $|t| \sim 1/L$, there are large sample to sample
fluctuations. As an illustration in the inset a) to Fig.\ref{FIG03}
we have shown the distribution of the finite-size specific heat,
$C^1_v$, defined for a given sample as the ratio of the distance
between the first two energy steps and the corresponding temperature
difference, just at the right of the transition point. The
distribution, as shown in the inset has a broad, power-law tail
and its second moment does not exist. This fact represents the
strong randomness character of the transition. Using continuous
distribution of disorder the strong fluctuation regime close to the
transition point remains qualitatively the same. Therefore, to determine
the thermodynamic singularity of the specific heat we analyzed its
behavior for $|t|>1/L$. As presented in the inset b) to Fig.\ref{FIG03}
the specific heat has a logarithmic singularity of the form of
$C_v(t) \sim (ln|t|)^{\epsilon}$, with $\epsilon \le 1$.
Thus the specific heat exponent is $\alpha=0$ and
the correlation length exponent is $\nu=1$. This latter result is the half
of the similar exponent of the RTIM in the strongly anisotropic IRFP\cite{fisher}.

%%%%%%%%%%%%%%%%%%%%%%%%%%%%%%%% FIG. 3 %%%%%%%%%%%%%%%%%%%%%%%%%%%%%%%%%%%%%%
%%%%%%%%%%%%%%%%%%%%%%%%%%%%%%%%%%%%%%%%%%%%%%%%%%%%%%%%%%%%%%%%%%%%%%%%%%%%%

To conclude our investigations have shown that the critical behavior
of the RBPM in the large-$q$ limit is dominated by strong disorder
effects and possibly related to the IRFP of the RTIM. The
conjectured exact values of the critical exponents are checked by
extensive numerical calculations. If an asymptotically exact
RG treatment - in the same spirit as the SDRG for the RTIM  - can be
constructed, should be clarified by future research.

We are indebted to R. Juh\'asz,
M. Preissmann, H. Rieger, A. Seb\H o and L. Turban for stimulating discussions.
This work has been supported by the Hungarian National
Research Fund under  grant No OTKA TO34183, TO37323,
MO28418 and M36803, by the Ministry of Education under grant No FKFP 87/2001,
by the EC Centre of Excellence (No. ICA1-CT-2000-70029).

\end{multicols}

\begin{references}

\bibitem{mccoywu}
        B.M. McCoy and T.T. Wu, 
        Phys. Rev. {\bf 176}, 631 (1968); {\bf 188}, 982(1969);
        B.M. McCoy, Phys. Rev. {\bf 188}, 1014 (1969).

\bibitem{2drg}
	O. Motrunich, S.-C. Mau, D.A. Huse and D.S. Fisher, Phys. Rev. B{\bf 61}, 1160 (2000).

\bibitem{fisher}
        D.S. Fisher, Phys. Rev. Lett. {\bf 69}, 534 (1992); 
        Phys. Rev. B {\bf 51}, 6411 (1995).

\bibitem{kogut}
	J. Kogut, Rev. Mod. Phys. {\bf 51}, 659 (1979).

\bibitem{profile}
	F. Igl\'oi and H. Rieger, Phys. Rev. Lett. {\bf 78}, 2473 (1997).

\bibitem{cardy}
	For a review see: J.L. Cardy, in {\it Phase Transitions and Critical Phenomena},
	edited by C. Domb and J.L. Lebowitz (Academic Press, London, 1987), Vol. 11.

\bibitem{Wu}
  F.Y. Wu, Rev. Mod. Phys. {\bf 54}, 235 (1982).

\bibitem{JRI01}
  R. Juh\'asz, H. Rieger, and F. Igl\'oi, Phys. Rev. E{\bf 64}, 056122 (2001).
 
\bibitem{aips02}
	J.-Ch. Angl\`es d'Auriac, F. Igl\'oi, M. Preissmann, and A. Seb\H{o},
	J. Phys. A{\bf 35}, 6973 (2002).

\bibitem{aizenmanwehr}
        M. Aizenman and J. Wehr, Phys. Rev. Lett. {\bf 62}, 2503
        (1989); errata {\bf 64}, 1311 (1990).

\bibitem{pottsmc}
        M. Picco, Phys. Rev. Lett. {\bf 79}, 2998 (1997); C. Chatelain and B. Berche,
        Phys. Rev. Lett. {\bf 80}, 1670 (1998); Phys. Rev. E{\bf 58} R6899 (1998);
        {\bf 60}, 3853 (1999); T. Olson and A.P. Young, Phys. Rev. B{\bf 60}, 3428 (1999).

\bibitem{pottstm}
        J.L. Cardy and J.L. Jacobsen, Phys. Rev. Lett. {\bf 79}, 4063 (1997),
        J.L. Jacobsen and J.L. Cardy, Nucl. Phys. B{\bf 515}, 701 (1998).       

\bibitem{dom_kinz}
        W. Kinzel and E. Domany, Phys. Rev. B{\bf 23}, 3421 (1981).


\bibitem{jacobsenpicco}
        J.L. Jacobsen and M. Picco, Phys. Rev. E{\bf 61}, R13 (2000);
        M. Picco (unpublished).

\bibitem{ccfs}
        J. T. Chayes, L. Chayes, D. S. Fisher and T. Spencer,
        Phys. Rev. Lett. {\bf 57}, 299 (1986).

\bibitem{staufferaharony}
  For a review see, Stauffer and A. Aharony, {\it Introduction to Percolation Theory},
  (Taylor and Francis, London) (1992).

\bibitem{mdh}
	S.K. Ma, C. Dasgupta and C.-K. Hu, Phys. Rev. Lett. {\bf 43}, 1434 (1979);
	C. Dasgupta and S.K. Ma, Phys. Rev. B{\bf 22}, 1305 (1980).

\bibitem{RFIM}
  J. C. Angl\`es d'Auriac and Nicolas Sourlas.\\
  Europhys. Lett. {\bf 39}, 473 (1997).


\end{references}
\end{document}